\documentclass[dvips]{article}
\usepackage{supertabular,lscape,epsfig}
\usepackage{amssymb}
\usepackage{amsmath}

\usepackage[polish]{babel}
\usepackage[T1]{fontenc}
\usepackage[latin2]{inputenc}
\usepackage{pslatex}

\DeclareSymbolFont{ppa}{OT1}{ppl}{m}{it}
\DeclareMathSymbol{\vv}{\mathalpha}{ppa}{'166}

\begin{document}

\newcommand{\dd}{\,{\rm d}}
\newcommand{\ie}{{\it i.e.},\,}
\newcommand{\etal}{{\it et al.\ }}
\newcommand{\eg}{{\it e.g.},\,}
\newcommand{\cf}{{\it cf.\ }}
\newcommand{\vs}{{\it vs.\ }}
\newcommand{\zdot}{\makebox[0pt][l]{.}}
\newcommand{\up}[1]{\ifmmode^{\rm #1}\else$^{\rm #1}$\fi}
\newcommand{\dn}[1]{\ifmmode_{\rm #1}\else$_{\rm #1}$\fi}
\newcommand{\upd}{\up{d}}
\newcommand{\uph}{\up{h}}
\newcommand{\upm}{\up{m}}  
\newcommand{\ups}{\up{s}}
\newcommand{\arcd}{\ifmmode^{\circ}\else$^{\circ}$\fi}
\newcommand{\arcm}{\ifmmode{'}\else$'$\fi}
\newcommand{\arcs}{\ifmmode{''}\else$''$\fi}
\newcommand{\MS}{{\rm M}\ifmmode_{\odot}\else$_{\odot}$\fi}
\newcommand{\RS}{{\rm R}\ifmmode_{\odot}\else$_{\odot}$\fi}
\newcommand{\LS}{{\rm L}\ifmmode_{\odot}\else$_{\odot}$\fi}

\newcommand{\Abstract}[2]{{\footnotesize\begin{center}ABSTRACT\end{center}
\vspace{1mm}\par#1\par   
\noindent
{~}{\it #2}}}

\newcommand{\TabCap}[2]{\begin{center}\parbox[t]{#1}{\begin{center}
  \small {\spaceskip 2pt plus 1pt minus 1pt T a b l e}
  \refstepcounter{table}\thetable \\[2mm]
  \footnotesize #2 \end{center}}\end{center}}

\newcommand{\TableSep}[2]{\begin{table}[p]\vspace{#1}
\TabCap{#2}\end{table}}

\newcommand{\FigCap}[1]{\footnotesize\par\noindent Fig.\  %
  \refstepcounter{figure}\thefigure. #1\par}

\newcommand{\TableFont}{\footnotesize}
\newcommand{\TableFontIt}{\ttit}
\newcommand{\SetTableFont}[1]{\renewcommand{\TableFont}{#1}}

\newcommand{\MakeTable}[4]{\begin{table}[htb]\TabCap{#2}{#3}
  \begin{center} \TableFont \begin{tabular}{#1} #4
  \end{tabular}\end{center}\end{table}}

\newcommand{\MakeTableSep}[4]{\begin{table}[p]\TabCap{#2}{#3}
  \begin{center} \TableFont \begin{tabular}{#1} #4
  \end{tabular}\end{center}\end{table}}

\newenvironment{references}%
{
\footnotesize \frenchspacing
\renewcommand{\thesection}{}
\renewcommand{\in}{{\rm in }}
\renewcommand{\AA}{Astron.\ Astrophys.}
\newcommand{\AAS}{Astron.~Astrophys.~Suppl.~Ser.}
\newcommand{\ApJ}{Astrophys.\ J.}
\newcommand{\ApJS}{Astrophys.\ J.~Suppl.~Ser.}
\newcommand{\ApJL}{Astrophys.\ J.~Letters}
\newcommand{\AJ}{Astron.\ J.}
\newcommand{\IBVS}{IBVS}
\newcommand{\PASP}{P.A.S.P.}
\newcommand{\Acta}{Acta Astron.}
\newcommand{\MNRAS}{MNRAS}
\renewcommand{\and}{{\rm and }}
\section{{\rm REFERENCES}}
\sloppy \hyphenpenalty10000
\begin{list}{}{\leftmargin1cm\listparindent-1cm
\itemindent\listparindent\parsep0pt\itemsep0pt}}%
{\end{list}\vspace{2mm}}
 
\def\TYLDA{~}
\newlength{\DW}
\settowidth{\DW}{0}
\newcommand{\dw}{\hspace{\DW}}

\newcommand{\refitem}[5]{\item[]{#1} #2%
\def\REFARG{#3}\ifx\REFARG\TYLDA\else, {\it#3}\fi
\def\REFARG{#4}\ifx\REFARG\TYLDA\else, {\bf#4}\fi
\def\REFARG{#5}\ifx\REFARG\TYLDA\else, {#5}\fi.}

\newcommand{\Section}[1]{\section{#1}}
\newcommand{\Subsection}[1]{\subsection{#1}}
\newcommand{\Acknow}[1]{\par\vspace{5mm}{\bf Acknowledgements.} #1}
\pagestyle{myheadings}

\newfont{\bb}{ptmbi8t at 12pt}
\newcommand{\xrule}{\rule{0pt}{2.5ex}}  
\newcommand{\xxrule}{\rule[-1.8ex]{0pt}{4.5ex}}  
\def\thefootnote{\fnsymbol{footnote}}
\begin{center}

{\Large\bf
The Optical Gravitational Lensing Experiment.\\
\vskip2pt
Final Reductions of the OGLE-III Data\footnote{Based on observations
obtained with the 1.3 m Warsaw telescope at the Las Campanas Observatory
of the Carnegie Institution of Washington.}}
\vskip1.5cm
{\bf A.~~U~d~a~l~s~k~i,~~ M.\,K.~~ S~z~y~m~a~ń~s~k~i,\\
  I.~~S~o~s~z~y~ń~s~k~i~~ and~~ R.~~P~o~l~e~s~k~i}
\vskip5mm
  Warsaw University Observatory, Al. Ujazdowskie~4, 00-478~Warszawa, Poland\\
e-mail: (udalski,msz,soszynsk,rpoleski)@astrouw.edu.pl
\end{center}

\Abstract{We describe methods applied to the final photometric
reductions and calibrations to the standard system of the images
collected during the third phase of the Optical Gravitational Lensing
Experiment survey -- OGLE-III. Astrometric reduction methods are
also presented.

The OGLE-III data constitute a unique data set covering the Magellanic
Clouds, Galactic bulge and Galactic disk fields monitored regularly
every clear night since 2001 and  being significant extension and
continuation of the earlier OGLE observations. With the earlier OGLE-II
and OGLE-I photometry some of the observed fields have now 16-year long
photometric coverage.}{Surveys -- Techniques: image processing --
Techniques: photometric}

\Section{Introduction} 
During the third phase of the Optical Gravitational Lensing Experiment
(OGLE-III), the observing capabilities of the OGLE project increased by
about order of magnitude compared to the previous OGLE-II phase
(Udalski, Kubiak and Szymański 1997, Szymański 2005) allowing
significant extension of the sky coverage. The OGLE-III phase started on
June 12, 2001 and has been continued up to now. Each clear night about 100
images ($>3$~TB raw data per year) are collected with the 1.3-m Warsaw
telescope equipped with eight CCD detector mosaic camera at Las Campanas
Observatory, Chile (operated by the Carnegie Institution of Washington).
The most important targets observed include the Magellanic Clouds,
Galactic center and Galactic disk fields. The main survey is conducted
for variable objects so most of observations are obtained in the {\it
I}-band filter.  Nevertheless, all fields are also observed from time to
time in the {\it V}-band with frequency of about 10\% of {\it I}-band
coverage.

The observing material already collected during the OGLE-III phase is a
unique dataset that can be used in a large variety of astrophysical
applications. The OGLE images are obtained at one of the best observing
astronomical sites worldwide so their quality -- seeing, background etc.
-- is usually very good. However, to fully utilize their quality the
state-of-the-art reduction methods must be applied. Also, the data should
be precisely calibrated so they could be compared with other data sets or
directly applied to astrophysical problems.

In this paper we describe reduction methods applied to the collected
OGLE-III images. The main goal is to obtain the most precise, well
calibrated photometry, as well as astrometry from the OGLE-III images.
As the OGLE-III data will eventually be publicly available this paper
can serve as a guide on the applied reduction procedures.

\Section{Photometry}
It was clear from the very beginning of the OGLE-III survey that the
photometric reductions of the huge number of images collected during this
phase should be performed in two steps: provisional on-line reductions at the
telescope implemented as soon as possible after starting OGLE-III and the
final reductions of the entire dataset.  Because of extreme stellar density
of the vast majority of fields observed by OGLE it was decided to use the
{\sc Difference Image Analysis} (DIA) method (Alard and Lupton 1998, Alard
1999, Woźniak 2000) as the primary photometric technique used for the
determination of photometry of hundreds of millions stars observed by OGLE.
 
The first step was implemented practically immediately after the start
of the OGLE-III phase -- in 2002. Because the DIA technique requires a
deep, good quality reference image (usually an average of several best
individual images) that serves as a template for all the remaining
images it should have the possibly best resolution (lowest possible
seeing) and low background. On the other hand the limited dataset of
collected images after the first one/two observing seasons and urgent
requirement of implementation of on-line reductions allowed construction
of reference images of only moderate quality, sometimes quite far from
the optimal images that could be collected with the OGLE-III hardware.
Thus, the on-line OGLE-III photometry, while providing fast and still
very precise magnitudes, was certainly not optimal from the photometric
point of view. The OGLE-III real-time data pipeline is described in
Udalski (2003).

After seven observing seasons the number of collected images during the
OGLE-III phase reached almost 200\,000 enough to start the huge project
of re-reductions of the whole OGLE-III observing material collected so
far with the goal of obtaining optimal and precisely calibrated final
OGLE-III photometry.

\subsection{Construction of Reference Images}

The preliminary reference images used for provisional OGLE-III on-line
photometry covered only the area corresponding to a single CCD detector
size ($2048\times4096$, 0\zdot\arcs26/pixel). Objects located in the
gaps between neighboring detectors in the OGLE-III mosaic camera were in
this way omitted. Because of imperfections in the telescope pointing the
gap regions were, however, imaged from time to time. Therefore for the
final reductions we decided to use extended size of the reference
images: of the total size of $2180\times4176$~pixels, so the gaps were
covered.

To construct the reference images for each field observed during
OGLE-III phase a subset of good quality images for each of the subfields
(corresponding to each CCD mosaic detector) of a given field was
prepared.  The selection of the best images was divided into two steps
-- automatic  and manual. In the first step all available images of a
given subfield  were classified using various criteria: seeing, sky
background level,  roundness of stellar profiles and ``cloudiness
indicator''. The latter  parameter was based on the existing provisional
photometry. It can be  expected that non-photometric or just cloudy
nights should result in  enormously large scatter of the measurements
obtained during these nights.  Thus, for every star brighter than
$I=18.0$~mag the mean magnitude and the  standard deviation were
derived, and then, for each point of each star the  deviation from the
mean magnitude was calculated and normalized by the  standard deviation.
Such coefficients were averaged over all objects in a  given subfield
to derive the final ``cloudiness indicators''.

Only images with the seeing smaller than $1\arcs$ ($1\zdot\arcs25$ for
$V$ band), the sky  background smaller than 1000 ADU (300 for $V$ band),
the ``roundness  parameter'' (defined as $1-b/a$, where $a$ and $b$ are
the largest and smallest diameter of stellar profile, respectively)
smaller than 0.1  and the ``cloudiness indicator'' smaller than 1.0 were
chosen to the second step of the selection procedure. Then images were 
arranged according to the seeing value and they were visually inspected to 
reject observations with unusually bad shapes of stellar profiles, with
trails of satellites, meteors or similar defects.

Up to 30 images for every subfield were selected, although every pixel
of  the final reference image consisted of maximum 10 the best available
observations. This excess of the images was essential for preparing
wide margins of the reference images.

After the standard procedure of aligning  the individual subset images
in the pixel grid, resampling fluxes using splines and rescaling fluxes
to the scale of the first image on the subset list, each pixel of the
reference image became an average of 10 or less, if not available,
corresponding pixels of the best subset images. The first image on the
subset list also defined the astrometric grid of the reference image.

Depending on the stellar density the photometric reductions were
performed on subframes into which a subfield was divided. The most dense
bulge fields in the {\it I}-band were divided into $545\times522$~pixel
subframes, while much less dense {\it V}-band images into
$1090\times1044$~pixel subframes. The Magellanic Cloud fields were
divided either into $1090\times1044$~pixel subframes or into
$2180\times2088$~pixel chunks for the less dense subfields.

\subsection{Photometry}
After construction of reference images all collected frames of a
subfield were processed using the photometric data pipeline similar to
the pipeline used in on-line reductions (Udalski 2003). The software was
adjusted to handle different size of subframes.

The PSF photometry on each DIA difference image was derived for each
object from the list of stellar objects on the reference image at its
position on the latter. The fitting radius was set to 5 pixels. The PSF
shape was derived during the transformation between a current subframe
and the reference image subframe of the processed subfield in the DIA
image processing. The difference flux was added to the reference image
base flux of the corresponding stellar object obtained with identical
photometry parameters and the total flux was converted to instrumental
magnitude. Additionally the difference images were subject of the search
for ``new'' stellar objects not associated with reference image stars.
This step secured the detection of moving objects or transient objects
brightening from below the detection threshold.

Each of the subframes after aligning with the corresponding reference
image subframe (\ie before image subtraction) was also processed with
the {\sc DoPHOT} photometry program (Schechter, Saha and Mateo 1993)
running in the standard ``non-fixed position'' mode. In this way
independent PSF profile photometry of each object was derived,
supplemented with  astrometric information (current $X,Y$ position)
obtained {\it via} PSF fitting. The zero point of the PSF photometry was
adjusted to the zero point of the DIA photometry of a given subframe by
comparison of appropriate DIA and PSF magnitudes of several dozen
brightest stars in each subframe.

Finally, the files containing the DIA and PSF photometry for each
subframe as well as files containing ``new'' objects were merged into
the final photometry file set for a given subfield image.

\subsection{OGLE-III Photometric Databases}
After processing the entire dataset of images for a given field the
instrumental photometry databases enabling easy handling of so huge amount
of data were constructed. The same databases as in the previous phases of
the OGLE survey were used. For technical details the reader is referred to
Szymański (2005) and Szymański and Udalski (1993). Because the OGLE-III
dataset of images is still open and the databases will be supplemented with
photometry from further observing seasons the standard ``integer'' data
format was used. Nevertheless all OGLE-III databases will be converted to
much faster ``sequential'' format when the OGLE-III phase ends.

The DIA photometric databases are constructed separately for each of the
subfields of a given field. They contain the following information for each
object detected in the reference image: epoch -- Heliocentric Julian Date
of the mid-exposure, magnitude, magnitude error from the DIA reductions,
photometry flags (\eg the detection of object on difference image for fast
identification as possible variable), position on the reference image. The
database of ``new'' objects contains similar information. However, the
brightness and its error for these objects are kept in flux units.

Additional PSF reduction procedure described in the previous Section
enabled construction of independent PSF photometric databases. Although the
PSF photometry is generally more noisy in the dense stellar fields it may
be necessary for correct identification of variable objects and in some
cases for more precise photometric calibration of the DIA photometry.

The PSF databases were also constructed separately for each subfield and
their structure is similar to the DIA photometry database. However, each
record for a given epoch, in addition to the photometric data
(magnitude, PSF magnitude error), contains also $(X,Y)$ position of the
object on this frame. In this way the PSF databases enable fast
retrieving of not only photometric but also astrometric data of a given
object (\cf Section~3.1). Because the detection level of stars processed
by the PSF photometry pipeline strongly depends on the seeing of a frame
the number of detected stars varies from image to image of the same
field. Thus, the cross-identification of detected objects is a necessary
step before feeding the PSF database records. Fortunately, the pixel
grid of all reduced images is the same (they are aligned with the DIA
reference image before processing), so the cross-identification is
straightforward: objects located within 1.9~pixel (0\zdot\arcs5) radius
from the reference image position are treated as associated with a given
reference image stellar object. One has, however, be aware that some
high proper motion stars can leave such an aperture after some time.

\subsection{Calibration}
During the first couple of years of the OGLE-III project the collected
images were reduced photometrically in the real time at the
telescope and the photometry was immediately fed into databases for
easy data handling. This provisional photometry was only roughly
calibrated based on very approximate zero points derived by comparison
of stellar magnitudes in OGLE-II fields observed in the course of the
OGLE-III survey. Accuracy of zero points of photometry was only about
$\pm0.1{-}0.2$~mag. Although that accuracy of absolute calibration was
certainly not sufficient for many projects, these data were fully useful
for all applications studying stellar variability.

Huge amount of data, large field of the sky covered by the images and very
dense stellar fields observed by OGLE make the standard system precise
calibration a difficult task. Here, we describe our methods of calibrating
OGLE-III photometry based on calibrations of the Magellanic Cloud
images. At the moment of writing this paper, the Galactic bulge data are
still being reprocessed. However the procedures that will be applied to
this latter data set will be practically identical.
\newpage

{\it Photometric Flatness of OGLE-III Images} 
\vspace*{3pt}

The first step in the procedure of calibrating the OGLE-III data was the
determination of the photometric flatness of each of the mosaic camera
images. Systematic spatial trends were encountered during calibration
of OGLE-II data and one could expect similar scale or larger effects
present in the OGLE-III images. This could be an effect of imperfections
in flat fielding, changing scale in the camera field and other
systematic effects.

To derive the maps of the flatness we conducted a dedicated observing
program carried out on 12 photometric nights during the years
2007--2008. We observed the Landolt's standard fields (PG2213-006 and
RUBIN~149; Landolt 1992) in nine positions on each chip: left, middle
and right in {\it X}-axis and bottom, middle and top in the {\it
Y}-axis. To ensure accurate measurements the seeing could not exceed 7
pixels. Measurements were obtained in both {\it V} and {\it I}-band
filters.

\begin{figure}[htb]
\centerline{\includegraphics[bb=60 60 380 250, width=11.5cm]{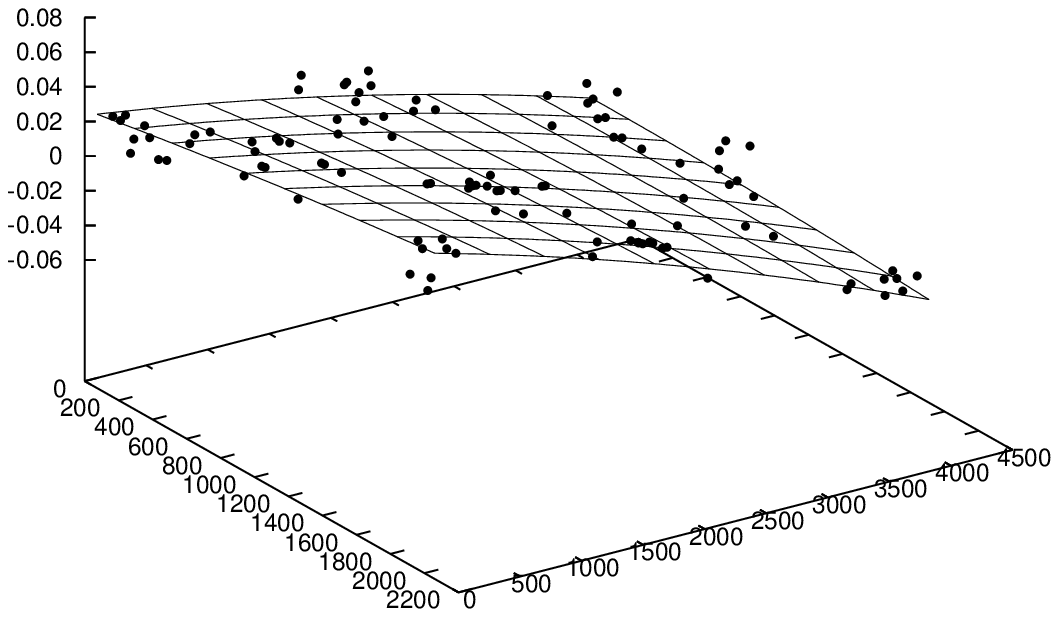}}
\FigCap{Flatness map derived from of observations of standard stars for
OGLE-III subfield \#2.}
\end{figure}
After collecting the data, magnitudes of standard stars were measured
with the aperture photometry with the radius of 14 pixels. Then we
normalized the magnitudes of each star taking as the zero point
magnitude of this standard star located in the center of the image for a
given observing dataset. In this way we obtained a map of the measured
magnitudes relatively to the chip center. Such maps were constructed for
each chip and each observing night. Finally we fitted a surface
$\Delta{\rm mag}=f(X,Y)$ to all individual maps for each chip. The
residuals of the typical surface fit were of the order of 0.015~mag.

As one could expect there are systematic changes of photometric signal
across the detectors. They are, however, very stable and can be accurately
mapped by our fitted surfaces. Fig.~1 shows the flatness map for one of
the OGLE-III mosaic chips.

{\it Observations of Photometric Standard Fields}
\vspace*{3pt}

OGLE-II phase photometric data were calibrated to the standard system using
the Landolt's (1992) standard fields observed on several dozen photometric
nights. To ensure homogeneity we adopted similar strategy in OGLE-III
phase and the same standard fields were supposed to be used as the main
calibrators. However, it soon turned out that the calibration of each
individual chip of the OGLE-III mosaic camera on many photometric nights
would be an extremely time consuming task limiting the time available for
scientific observations. As a result the standard fields were observed
relatively rarely and often the standards were observed on single
chip only. Nevertheless, during the course of the OGLE-III phase the Landolt
fields were observed on more than 30 photometric nights during the years
2003--2008 allowing determination of accurate transformation of the
instrumental magnitudes to the standard system. Calibration of photometry
of each particular chip to the standard system could be derived on more
than 20 photometric nights.

All standard stars in the observed fields were measured using aperture
photometry with the aperture radius of 14 pixels. Then the magnitudes were
corrected for the photometric flatness of respective chip using maps
described above. Next, for each chip, the mean nightly transformations
in the form:
$$V=\vv+k_V\cdot X+{\epsilon_V}\cdot(V-I)+ZP_V$$
$$I=i  +k_I\cdot X+{\epsilon_I}\cdot(V-I)+ZP_I$$ 
were derived, where lowercase letters describe aperture instrumental
magnitudes normalized to one second exposure time, $k$ denotes the
extinction coefficient, $X$ -- air mass, $\epsilon$ -- color term coefficient
and $ZP$ -- the zero point of the transformation.

\MakeTable{ccc}{12.5cm}{Color term coefficients for the OGLE-III subfields}
{\hline
Subfield &  $\epsilon_V$ &   $\epsilon_I$\\
\hline
Chip~1 &  $-0.039 \pm 0.010$ &   $0.050 \pm 0.016$\\
Chip~2 &  $-0.038 \pm 0.008$ &   $0.040 \pm 0.017$\\
Chip~3 &  $-0.036 \pm 0.010$ &   $0.042 \pm 0.019$\\
Chip~4 &  $-0.045 \pm 0.010$ &   $0.044 \pm 0.015$\\
Chip~5 &  $-0.040 \pm 0.009$ &   $0.043 \pm 0.011$\\
Chip~6 &  $-0.039 \pm 0.011$ &   $0.040 \pm 0.015$\\
Chip~7 &  $-0.037 \pm 0.014$ &   $0.039 \pm 0.015$\\
Chip~8 &  $-0.043 \pm 0.007$ &   $0.044 \pm 0.012$\\
\hline}

It turned out that the color term transformation coefficients behaved
quite stable over the several years long observations during the
OGLE-III phase. Therefore we decided to average the derived values of
these coefficients for the future use with the final databases. Table~1
lists the adopted $\epsilon$ coefficients for each chip.

\vspace*{6pt}
{\it Calibration of the OGLE-III Fields with Standard Stars}
\vspace*{3pt}

Only a small fraction of the OGLE-III fields could be calibrated directly
with the observations of standard stars, because of limited number of
nights when the standard fields were observed. Nevertheless we performed
such a calibration to compare calibrated magnitudes with independent data
set, namely OGLE-II photometry, and in this way we could double check
consistency of the applied procedures.

Calibration of the OGLE-III instrumental photometry to the standard
system with observations of the standard stars was performed in the
following steps. First, the aperture corrections to instrumental
magnitudes from the DIA data pipeline were derived. Because of high
stellar density of the vast majority of the OGLE-III fields a special
procedure was developed. Each of the subframes on which the DIA
photometry had been derived was analyzed independently. On each of them
7--170 brightest stars (in the Magellanic Cloud fields) with no
companions of comparable brightness within the radius of 20 pixels were
selected as stars for aperture correction determination. Then the
customary program based on the {\sc DoPHOT} photometry program was run
on each of the subframes. First, it derived the PSF photometry of all
objects on the subframe. Then it subtracted from the original subframe
all the stars except the ones on the aperture correction list,
minimizing in this way contamination of aperture stars by fainter
neighbors. Finally it calculated the aperture correction for each star
from the aperture correction star list by subtracting its aperture
magnitude measured in 14 pixel aperture on so prepared cleaned image
from the DIA magnitude obtained with the pipeline reductions. The median
value of the aperture corrections from all selected stars was taken as
the final aperture correction of a given subframe. Aperture corrections
were calculated for all OGLE-III fields observed on ``standard stars''
nights.

Then, the offset between the DIA instrumental magnitudes and standard
system was calculated for each subframe for a given ``standard stars''
night. The offset included the aperture correction, one second exposure
normalization, extinction term, $k\cdot X$, and the zero point value for
a given night (as obtained from standard stars observations). The
average of these offsets from all standard nights was then used as the
calibration shift between OGLE-III instrumental and standard systems.
Such a correction does not include the color term, $\epsilon\cdot
(V-I)$, which is individual for each object.

\vspace*{6pt}
{\it Comparison with OGLE-II Photometry}
\vspace*{3pt}

To compare the standard stars calibration of the OGLE-III photometry
with independently calibrated OGLE-II dataset we cross-identified all
OGLE-II stars with OGLE-III counterparts based on astrometric solutions
(see Section~3.2). Only OGLE-III subfields (corresponding to an
individual detector image of a given OGLE-III field) with overlap larger
than 75\% with OGLE-II fields were used in this test. Also, for
comparisons only the brighter stars were selected: $V<19$~mag and
$I<18$~mag, with standard deviation of all measurements smaller than
0.04~mag.

78 and 53 subfields in the LMC were suitable for comparison of the
OGLE-III and OGLE-II photometry in the {\it I} and {\it V} bands,
respectively (they fulfilled overlap criteria and were calibrated with
standard stars). Unfortunately none of the OGLE-III SMC fields that
overlap with OGLE-II were observed in the {\it V} filter on the
``standard stars'' nights, so in the SMC only {\it I}-band photometry
could be compared (48 subfields). Altogether 622\,272 LMC stars were
used for comparison of OGLE-III standard star calibration with OGLE-II
photometry in the {\it I}-band and 698\,847 in the {\it V}-band. In the
SMC 117\,743 stars were compared in the {\it I}-band.

OGLE-II calibrated magnitudes were taken from the LMC and SMC
photometric maps (Udalski \etal 1998, 2000). The OGLE-III instrumental
magnitudes were shifted using the calibration offsets described above,
corrected for photometric flatness and for the color term, using OGLE-II
$V-I$ color and average color coefficients (Table~1). The shifts between
the OGLE-III and OGLE-II magnitudes of all selected common stars were
then calculated. 

The median shift between the OGLE-III and OGLE-II magnitudes was
determined for each subframe of each analyzed subfield and the average
value for each subfield was calculated. The final difference of the
OGLE-III photometry calibrated directly with standard stars and OGLE-II
photometry, calculated as the mean of all comparison subfields in a
given OGLE-III target and filter, and its standard deviation is listed
in Table~2.

\MakeTable{lcc}{12.5cm}{Mean difference between OGLE-III
photometry calibrated directly with standard stars and OGLE-II photometry}
{\hline
LMC {\it I}-band: &   $-0.004 \pm 0.013$ &    71 subfields\\
LMC {\it V}-band: &   $-0.003 \pm 0.013$ &    53 subfields\\
&&\\
SMC {\it I}-band: &   $+0.004 \pm 0.013$ &    48 subfields\\
\hline}

As it can be seen from Table~2 the agreement between the OGLE-III and
OGLE-II magnitudes is extremely good. Practically the shifts between
both datasets are negligible what is reassuring and indicates that the
calibration procedures are self consistent. This also indicates that the
huge OGLE-II dataset can serve as the large set of secondary standard
stars enabling simple and easy way of calibration of all OGLE-III
fields. Practically on each night several OGLE-III fields that overlap
with OGLE-II ones have been observed during the OGLE-III survey
providing calibrating standards for the remaining OGLE-III fields.

Finally, the shifts between OGLE-III and OGLE-II  photometry  of
hundreds of thousands stars from all overlapping subfields were used for
the determination of second order spatial corrections to the first
approximation correction for photometric flatness obtained from
measurements of standard stars.  Huge sample of common stars distributed
uniformly over entire chips allowed more precise mapping of low level
trends than limited standard stars measurements. 

First, to minimize the possible low level systematic errors in the
OGLE-II maps (resulting from the reduction and calibration methods of
OGLE-II drift-scan images) the shifts of all common stars were  plotted
as a function of $X$ position of a star in the OGLE-II image. Then the
mean shift level as a function of OGLE-II $X$ coordinate was fitted with
splines and OGLE-II magnitudes were detrended (\ie shifts were
corrected; the correction was always in the range ($-0.014,0.010$)~mag).
Finally, the mean values of so adjusted shifts in appropriate pixels of
each OGLE-III mosaic chip were found and the chip surfaces were smoothed
forming second order photometric flatness correction maps of each mosaic
detector.  

\vspace*{6pt}
{\it Establishing the Set of OGLE-II Standards}
\vspace*{3pt}

Results of the comparison of OGLE-III standard stars calibrated photometry
with OGLE-II photometry imply that the OGLE-II photometry can be used as
the set of secondary standards for calibrating purposes. Therefore, for
the final calibrations of OGLE-III photometry we selected a set of OGLE-III
subfields that overlap with OGLE-II fields on more than 75\%. Altogether 92
such subfields from 23 OGLE-III fields in the LMC and 48 subfields from 11
OGLE-III SMC fields were selected as ``photometric calibrators''.

Before we started calibration procedures using OGLE-II fields, we prepared
lists of ten best images of all observed OGLE-III fields. In the {\it
I}-band we used the same criteria as in the selection of frames for
construction of the reference images (seeing < $1''$, background < 1000
ADU), but this time we sorted images according to the ``cloudiness
indicator'' and selected images with the lowest value of this parameter
(usually smaller than 0.6). Because of much smaller number of {\it V}-band
images and general observing strategy that {\it V}-band observations are
taken on ``photometrically looking'' nights all {\it V}-band observations
with seeing smaller than 1\zdot\arcs5 were considered as the best images.

All nights when the best images were taken, supplemented by the nights
when the photometric conditions were confirmed by observations of
standard stars, were assumed to be ``calibrating nights''. All
observations of ``photometric calibrators'' taken on these nights were
used for the determination of zero points for a given night. Standard
deviation of the zero points derived from different ``photometric
calibrators'' subfields served as the consistency check of the
photometric condition of a night. Only the nights when $\sigma_{\rm
ZP}<0.03$~mag were considered as photometric.

The OGLE-III DIA pipeline photometry of the ``photometric calibrators''
subfields was color term corrected based on the OGLE-II color index,
corrected with aperture correction derived for each subframe, extinction
corrected and compared with OGLE-II photometry. The median of the
magnitude shifts for the brightest stars (fulfilling similar magnitude
and error limitations as in the previous Subsection) was considered as
the photometric scale zero point. Independent determinations from
different calibrators were then averaged and their scatter served as the
measure of the photometric quality of the night.

Altogether about 180 photometric nights for each of the mosaic CCD
detectors calibrated using the OGLE-II LMC and SMC ``photometric
calibrators'' were used for calibration of the {\it I}-band images and
about 130 for {\it V}-band ones.

As another consistency check we compared the {\it I}-band zero points
derived with OGLE-II ``photometric calibrators'' with direct determinations
using standard stars described above. The mean difference of the zero
points obtained with the two methods was found to be: $-0.001\pm0.017$~mag
for 92 common determinations. This confirms full agreement of the OGLE-II
calibration approach with classical calibration method using standard
stars.

\vspace*{6pt}
{\it Final Calibration of the OGLE-III Photometry}
\vspace*{3pt}

With the determination of the zero points of OGLE-III photometry on so
many photometric nights using OGLE-II ``photometric calibrators'', it
became possible to calibrate all the remaining OGLE-III fields. First,
for all LMC and SMC images taken on ``calibrating nights'' the aperture
corrections were derived using procedure described previously. Then for
each subframe in a given subfield photometric offsets  as in the
calibration with standard stars were calculated using just derived zero
point for a given night. Finally, the average offsets from all nights
for each subframe were determined after removing two extreme (largest
and smallest) values in the {\it I}-band and one extreme value in the
much smaller {\it V}-band set. Typical number of individual offsets
averaged for a subfield was 64 and 34 for the SMC and LMC {\it I}-band
data and 22 and 12 for {\it V}-band observations, respectively. The
typical scatter of individual offsets around the mean value was about
0.010~mag.

\vspace*{6pt}
{\it Correction of Databases}
\vspace*{3pt}

In the final step the databases containing the OGLE-III photometry in
the instrumental system were adjusted in such a way that the final
OGLE-III database of a given subfield for a given band contains 
calibrated magnitudes of each object except for the color term,
$\epsilon\cdot(V-I)$, because no color information about the object is
{\it a priori} known. The correction to photometry of each object
included the flatness correction, the second order flatness correction
-- both dependent on the position of the star in a subfield -- and the
mean offset for a subframe of a given subfield. Because the PSF
instrumental photometry was tied during the reduction procedure to the
DIA photometry, the same corrections apply to the same object in the DIA
and PSF databases.

To calculate calibrated OGLE-III photometry of a given star one has to
obtain a color information on the object to correct database magnitudes for
the color term. If this object is present in the other band OGLE-III
database the procedure is straightforward: one should
extract and then calibrate the $(\vv-i)$ instrumental color according to
the following equation:
$$(V-I)=(\vv-i)/(1-\epsilon_V+\epsilon_I)$$
and then apply the color term correction using the transformation
color coefficients for a given chip provided in Table~1. When a star does
not have OGLE color information one can assume a typical color of the
population of stars in the analyzed sample. Accuracy will be
somewhat lower, but on the other hand the color term is a second order
correction so the inaccuracy of such an approximation should be small.

Figs.~2--9 present a few examples of the comparison of OGLE-III
calibrated photometry with the OGLE-II Photometric Maps for subfields
from the center and outer regions of the galaxy bar in the {\it I} and
{\it V}-band. One can see that the residual differences and/or some
residual trends are small, at most of the order of 0.01--0.02~mag, \ie
practically negligible. These small trends are mostly noticeable as a
function of  OGLE-III $Y$ coordinate, \ie the OGLE-II $X$ coordinate and
seem to be correlated with OGLE-II subframe division. Thus their origin
is likely related to small systematic effects in the OGLE-II
calibration.

\Section{Astrometry}
CCD images can be successfully applied to astrometric applications. Good
resolution quality of OGLE images obtained at Las Campanas Observatory --
one of the best observing sites worldwide -- makes this material very
attractive for astrometric applications. Indeed the OGLE-II images were
successfully used for construction of the proper motion catalogs of the
Magellanic Cloud and Galactic bulge line-of-sight objects (Soszyński
\etal 2002, Sumi \etal 2004). OGLE-III images are even more attractive
for astrometric purposes as the finer sampling of 0\zdot\arcs26/pixel is
better suited to the typical observing seeing conditions. Astrometry 
from OGLE-III should be then more precise than from OGLE-II frames.

\subsection{Astrometric Databases}

The PSF photometric databases described in Section~2.3 can also serve as
astrometric databases. They contain current ($X,Y$) position of each object
detected in the DIA reference image on every collected OGLE-III image. This
position is derived by PSF fitting ($X,Y$,flux) with the {\sc DoPHOT}
program after aligning the current image with the reference image and
transformation to the same pixel grid. Astrometric databases provide then
direct access to proper motion measurements of a given object in $X$ and
$Y$ coordinates. $X$ and $Y$ axes of the OGLE-III images are roughly
aligned with N-S and E-W directions, respectively.  Fig.~10 shows OGLE-III
data of an exemplary high proper motion object taken from Soszyński's \etal
(2002) catalog.

\setcounter{figure}{9}
\begin{figure}[htb]
\centerline{\includegraphics[bb=0 50 510 510, width=12.5cm]{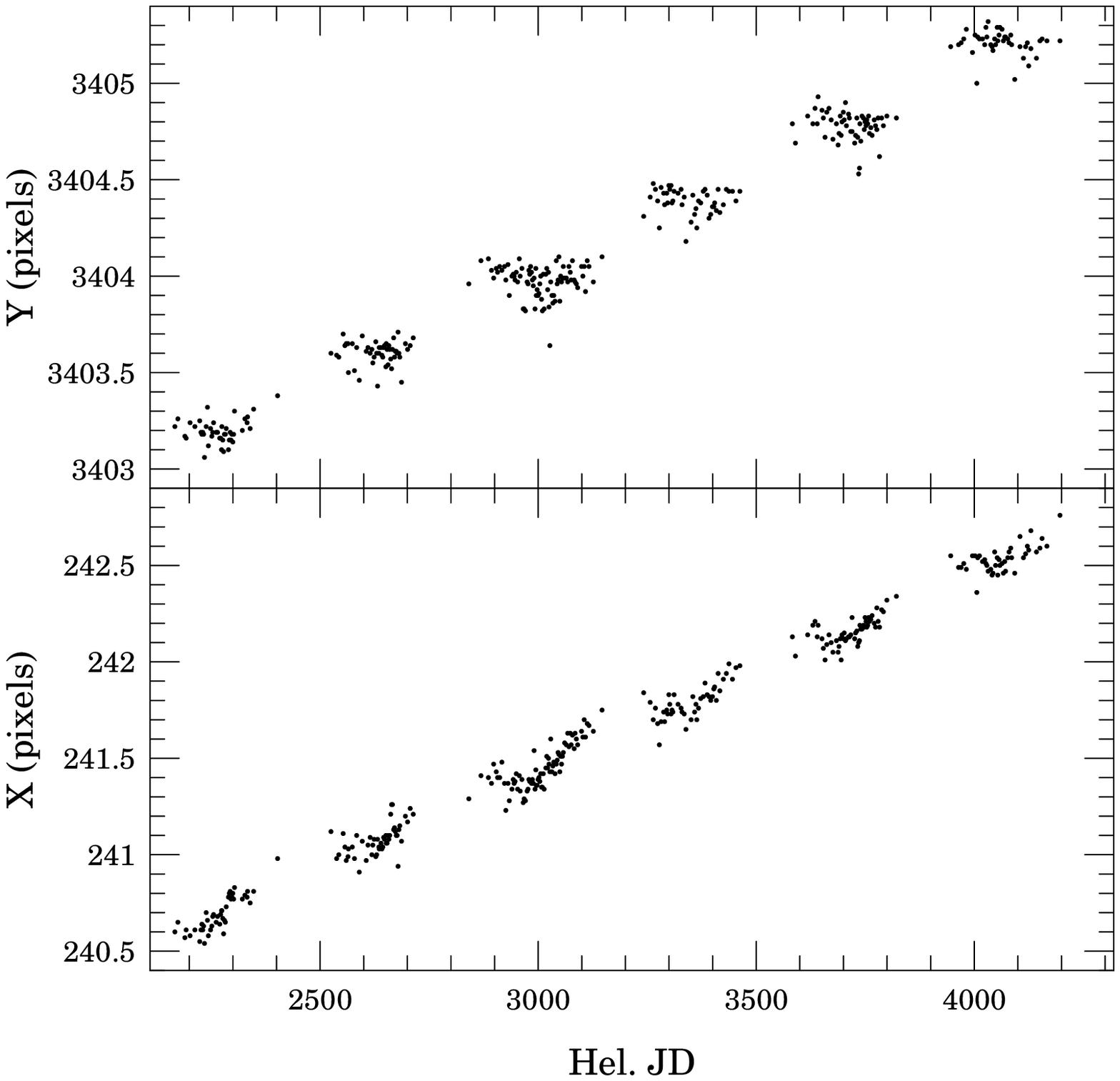}}
\FigCap{Astrometry of high proper motion object LMC\_SC13~351-352
(Soszyński \etal 2002). One pixel corresponds to 0\zdot\arcs26.}
\end{figure}

\subsection{Transformation of Pixel Coordinates to the Equatorial System}
The astrometric transformation of the OGLE-III pixel grid coordinates to
the equatorial system was performed in similar way as in the previous
phases of the OGLE project (Udalski \etal 1998) with some modifications
required to allow for much larger area covered by the OGLE-III
observations, including fields of significantly lower star density.

Each subfield of the OGLE-III targets, corresponding to a single chip of
the mosaic CCD camera had its astrometric transformation computed
independently. Then stars were cross-identified in the overlapping regions
of the neighboring reference images and their RA/DEC coordinates computed
from two independent transformations were compared as the test for the
accuracy of the astrometric calibration. Typical error estimated by this
comparison was about 0\zdot\arcs06. This test was made much simpler than in
former OGLE phases after introducing reference images which cover larger
area than the single chip of the CCD camera.

The procedure of finding the coordinates transformation for each subfield
was performed in following steps. As the preparation, the lists of
brightest stars in the subfield were extracted from the OGLE database and
from the chosen external catalog. A few catalogs were tested, including the
Second USNO CCD Astrograph Catalog (UCAC2, Zacharias \etal 2004), the
Catalog of Astrometric Standards USNO-A2.0 (Monet \etal 1998) and the Two
Micron All Sky Survey (2MASS) Point Source Catalog (Skrutskie \etal
2006). The best of these, for the purpose of finding our transformation,
proved to be the 2MASS Catalog and it was uniformly used for all fields in
the Magellanic Clouds.

In the second step, a short list of identified stars and a preliminary
transformation between the $X,Y$ pixel coordinates from the OGLE database
and the RA/DEC from the 2MASS Catalog list was found using the WCSTools
software package (Mink 2002).

In the last step, the third-degree polynomial transformation was calculated
using as many stars from both lists as we could cross-identify. The same
algorithm was used as in previous phases of the OGLE project, thus allowing
to add the results to the common OGLE database of astrometric
transformations and to use the same software. The transformations were
carefully tested for uniformity and accuracy, as explained above.

\Section{Summary}

OGLE-III databases constitute a unique dataset containing long term time
series photometry of hundreds of millions stars. The targets observed
during the OGLE project, namely the Magellanic Clouds and the Galactic
center are very important from the astrophysical point of view thus
making these databases especially attractive for data mining and
statistical analysis of large samples of objects of different types.

In this paper we presented technical details on processing the collected
images to derive the most precise and well calibrated photometry. Also
details on astrometric reductions were provided.

The finally reduced OGLE-III data will be gradually released to public
domain. Information on the available data sets can be found at:

\begin{center}
{\it http://ogle.astrouw.edu.pl}
\end{center}

\Acknow{This paper was  partially supported by the Polish MNiSW grants:
N20303032/4275 to AU and NN203293533 to IS and by the Foundation for
Polish Science through the Homing Program.

This publication makes use of data products from the Two Micron All Sky
Survey, which is a joint project of the University of Massachusetts and the
Infrared Processing and Analysis Center/California Institute of Technology,
funded by the National Aeronautics and Space Administration and the
National Science Foundation.}

\newpage
\centerline{Captions of JPEG figures.}
\vskip15pt
\noindent
Fig.~2. Difference between the {\it I}-band OGLE-III calibrated 
photometry and OGLE-II photometry for common stars from the central bar
subfield LMC100.1. {\it Upper panel} shows the difference as a function of
$X$ image coordinate (N-S direction) while the {\it lower panel} as a
function of $Y$ coordinate (E-W direction).
\vskip15pt
\noindent
Fig.~3. Same as Fig.~2 for the bar subfield LMC111.2.
\vskip15pt
\noindent
Fig.~4. Same as Fig.~2 for the western part of the bar subfield LMC126.1.
\vskip15pt
\noindent
Fig.~5. Same as Fig.~2 for the eastern part of the bar subfield LMC177.3.
\vskip15pt
\noindent
Fig.~6. Same as Fig.~2 for the {\it V}-band.
\vskip15pt
\noindent
Fig.~7. Same as Fig.~3 for the {\it V}-band.
\vskip15pt
\noindent
Fig.~8. Same as Fig.~4 for the {\it V}-band.
\vskip15pt
\noindent
Fig.~9. Same as Fig.~5 for the {\it V}-band.

\end{document}